\documentclass[conference]{IEEEtran}
\usepackage[normalem]{ulem}
\usepackage{cite}
\usepackage{amsmath,amssymb,amsfonts}
\usepackage{algorithm2e}
\usepackage{graphicx}
\usepackage{textcomp}
\usepackage[hyphens]{url}
\usepackage{comment}
\usepackage{soul}
\usepackage{ifthen}
\usepackage{calc}
\usepackage{pifont}
\usepackage{color}
\usepackage{fancyhdr}
\usepackage{xspace}
\usepackage{xkeyval}
\usepackage{multirow}
\usepackage[table]{xcolor} 
\usepackage{proof}
\usepackage{amsmath}
\usepackage{amsthm}
\usepackage{amscd}
\usepackage{amssymb}
\usepackage{wrapfig}

\newenvironment{stripetabular}{\rowcolors{2}{white}{lightgray}\tabular}{\endtabular}

\newcounter{hours}
\newcounter{minutes}

\newcommand{\name}{{QED}\xspace}

\newboolean{timeofmake}
\setboolean{timeofmake}{false}

\newcommand{\dontinclude}[1]{ }

\newcommand{\putsec}[2]{
\section{#2}\label{sec:#1}
}

\newcommand{\putsubsec}[2]{
\subsection{#2}\label{sec:#1}
}

\newcommand{\putsubsubsec}[2]
{
\subsubsection{#2}\label{sec:#1}\vspace{0.0in}}

\newcommand{\tabput}[3]{
\begin{scriptsize}
\begin{table}[t]
\vspace{-0.1in}
\caption{#3 \label{tab:#1}}
\begin{center}
{
#2
}
\vspace{-0.1in}
\end{center}
\end{table}
\end{scriptsize}
}

\newcommand{\figput}[4][1.0\linewidth]{
\begin{figure}[t]
\begin{minipage}{\linewidth}
\footnotesize 
\begin{center}
\includegraphics[width=#1]{figures/#2}
\end{center}
\vspace{-0.1in}
\caption{#4 \label{fig:#2}}
\end{minipage}
\vspace{-0.2 in}
\end{figure}
}

\newcommand{\figputW}[4][\linewidth]{
\begin{figure*}
\begin{minipage}{\linewidth}
\footnotesize 
\begin{center}
\includegraphics[width=#1]{figures/#2}
\end{center}
\vspace{-0.1in}
\caption{#4 \label{fig:#2}}
\vspace{-0.2in}
\end{minipage}
\end{figure*}
}

\newcommand{\figref}[1]{Figure~\ref{fig:#1}}
\newcommand{\tabref}[1]{Table~\ref{tab:#1}}
\newcommand{\secref}[1]{Section~\ref{sec:#1}}

\colorlet{highlightyellow}{yellow!50}
\sethlcolor{highlightyellow}
\newcommand{\lp}{$<_p\xspace$ }
\newcommand{\lm}{$<_m\xspace$ }
\newcommand{\lte}{$<_e\xspace$}
\newcommand{\lo}{$<_e\xspace$ }

\def\BibTeX{{\rm B\kern-.05em{\sc i\kern-.025em b}\kern-.08em
    T\kern-.1667em\lower.7ex\hbox{E}\kern-.125emX}}

\title{QED: Scalable Verification of Hardware Memory Consistency}
\author{
\IEEEauthorblockN{Gokulan Ravi, Xiaokang Qiu, Mithuna Thottethodi, T. N.Vijaykumar}
\IEEEauthorblockA{\textit{Elmore Family School of Electrical and Computer Engineering, Purdue University}\\
\{ravig,xkqiu,mithuna,vijay\}@purdue.edu}
}

\begin{document}
\maketitle
\thispagestyle{plain}
\pagestyle{plain}

\begin{abstract}

Memory consistency model (MCM) issues in  general-purpose, high-performance, out-of-order-issue
microprocessor-based shared-memory systems are notoriously non-intuitive and a source of hardware design bugs. Previous hardware verification work 
is limited to in-order-issue processors, 
to proving the correctness only of some test cases, or to bounded verification that does not scale in practice beyond 7 instructions across all threads.
Because cache coherence (i.e., write serialization and atomicity) and  pipeline front-end verification
and testing are well-studied, 
we focus on the memory ordering in an out-of-order-issue processor's load-store queue and the coherence interface between the core 
and global coherence. We propose QED
based on the key notion of {\em observability} that any hardware  reordering  matters only 
if a forbidden value is produced. We 
argue that one needs to consider (1) only  
{\em directly-ordered} instruction pairs -- transitively non-redundant pairs connected by an edge in the MCM-imposed partial  order --
and not all in-flight instructions, and (2)  only the ordering of external events from other cores  (e.g., invalidations) but not the events' originating cores, 
achieving verification scalability in  both the  numbers of 
in-flight memory instructions and of cores. 
Exhaustively considering all pairs of instruction types and all types of external events intervening between each pair, 
QED attempts to {\em restore} any reordered instructions to an MCM-complaint order without changing the execution values (i.e., unobservably), where failure indicates an MCM violation. 
Each instruction pair's exploration 
results in a decision tree of 
simple, narrowly-defined predicates to be  evaluated against the RTL implementation.
In our experiments, we automatically generate the decision trees for SC, TSO, and RISC-V WMO, and illustrate automatable verification by evaluating a substantial predicate against BOOM v3 implementation of RISC-V WMO, leaving full automation to future work.

\end{abstract}
\putsec{intro}{Introduction}
Memory consistency issues in  general-purpose, high-performance
microprocessor-based shared-memory systems are notoriously non-intuitive and complex. Memory 
consistency is a significant source of hardware design bugs~\cite{arm_bug, intel_bug, amd_bug}
which can lead to serious correctness issues, such as 
data corruption, incorrect lock behavior, and  crashes. 
While testing identifies some bugs, exhaustive testing to guarantee correctness is unrealistic.  As such,
the only viable option to  guarantee correct behavior at any system scale is verifying the implementation against the memory consistency model (MCM). 
However, out-of-order memory accesses and multiple levels of buffering and caching
introduce a myriad of interactions affecting memory ordering~\cite{tutorial},
making verification  profoundly challenging. 

The central issue is that the verification method must scale with the system size (e.g., the number of in-flight memory accesses in a core). Otherwise, verification would require examining  a  number of cases that  explodes combinatorially with the system size. Such intractability -- common in verification -- would  mean incomplete, inconclusive verification that may not be useful in practice for real system scales (i.e., few correctness guarantees). 
Moreover, ideally, the proof should be rigorous, machine-generated or machine-assisted, and against concrete RTL implementations. 

Early ``*check'' papers~\cite{pipecheck,ccicheck,coatcheck,rtlcheck}
apply all interleavings of each of a few tens to hundreds of short test programs to  microarchitectures  (RTL implementations) or their specifications (e.g., Intel's  ``litmus tests'', each of which typically comprises 4-8 memory accesses in 2-4 threads). Though the method {\em correctly} catches {\em all} the bugs  exposed by these tests, there may be bugs not exposed by these tests and therefore not caught~\cite{rtlcheck,pipeproof,comprehensive_asplos17,compare_mcms}.
Pointing to this insufficiency, 
a later work~\cite{comprehensive_asplos17} generates
exhaustive yet minimal tests with a bounded number of instructions ($n$) across all threads. 
but does not scale in practice beyond $n$ = 7, far fewer than modern instruction window sizes (e.g., hundreds of instructions). 
Later ``*check'' papers~\cite{tricheck,ilamcm,realitycheck} (and retroactively, the earlier papers) can leverage the exhaustive tests to achieve bounded verification. 
Acknowledging the  test-based approach's limitations, PipeProof~\cite{pipeproof} replaces the tests with arbitrary instruction sequences. The paper formulates the problem as a SMT instance and exploits the transitivity of happens-before relationships among microarchitecture events. 
However, the approach explores increasingly longer instruction sequences which requires manual invariants (with proofs) to terminate.   
Further, PipeProof and  Kami~\cite{kami}, a rigorous, modular approach, verify only
in-order issue pipelines which are far simpler than modern out-of-order issue processors. Instead, we propose an approach for out-of-order issue processors independent of, and hence scalable in, the numbers of instructions and of threads (cores).
Finally, an alternative approach proposes additional hardware  to dynamically ensure MCM correctness \cite{verify_stage} which increases cost and requires the new hardware itself to be verified.
In contrast, we  target static verification with no hardware overhead. 

We propose {\em QED}, scalable verification of
memory consistency for modern out-of-order issue processors and memory systems. While previous  unbounded verification~\cite{pipeproof, kami} has considered entire, simple, in-order issue pipelines, proving the correctness of an  entire out-of-order issue processor is hard and may be intractable. However, many consistency bugs arise from reordering and 
overlapping of memory accesses by the load-store queue and the memory hierarchy~\cite{arm_bug,intel_bug}. In contrast, any design bugs in the pipeline front-end related to register dependencies  would likely result in  not only consistency failure  but also incorrect sequential execution, and would likely  be caught by verification~\cite{isaformal} targeting the front-end  components. As such, we assume that the front-end register and control-flow dependencies are  implemented correctly.
Further, while cache coherence affects consistency, there is much previous work~\cite{futurebus,verif_survey,verif_symmetry,state_enumeration,symbolic_sm,ssm_three_protocols,param_verif, fractal++, fractal, pvcoherence,pv_modelchecking,verif_aggregation,verif_refinement,verifiable_hierarchical,push_button_verif,verify_bcast_mcast,lamport_clocks,talupur2008going,krstic2005parameterized} on verifying cache coherence. As such, we assume that  write serialization and, if required by the MCM,  write atomicity have  been verified as part of standard coherence verification.
However, because {\em events external to the core} -- e.g., cache misses, invalidation acknowledgments, incoming invalidations, and incoming read requests -- affect consistency,
we consider these events in {\em the coherence interface} between global coherence and the node (specifically, the core's load-store queue and local cache hierarchy). 
As such, to remain tractable while capturing the most relevant issues, we focus on memory ordering in the load-store queue  and  the coherence interface, whose 
unbounded verification is challenging and not covered  by previous work (bounded verification of up to 7 instructions in practice~\cite{comprehensive_asplos17} can be combined retroactively with previous work~\cite{ccicheck,coatcheck}). 

QED makes the following contributions:
 
A key challenge in traditional verification approaches is the
state space explosion that results from naively modeling the hardware. The first of two key scalability issues is 
 the number of in-flight memory instructions in a core (e.g., $n=100$), 
which may be reordered arbitrarily. Rather than consider this large space (potentially $n!$ reorderings),
we argue that {\em only  
transitively non-redundant instruction  pairs connected by an edge in the MCM-imposed partial  order, called directly-ordered instruction pairs,}
and intervening external events, which are proxies for instructions in other cores  (e.g., incoming invalidations and read requests),  need to be considered for MCM compliance; 
and that among the events relevant to an instruction, only
one event per instruction needs to be considered at a time.
Informally, this pairwise consideration suffices because  any illegal reordering of instruction A must also reorder A past  an instruction B connected to A by an edge  in the MCM-compliant partial order.

The second key scalability issue is 
the number of cores in the system.
We observe that {\em while considering external events,  we need to examine only the ordering among a  core's incoming events and  instructions but not from which cores the events originate  because implementations do not consider the events' origins.} This observation allows QED to  capture actions by other cores {\em independent} of their number. 
Combining the first two contributions, we need to consider only all pairs of {\em types} of in-flight memory instructions ($m$ types) and {\em only one of the types} of intervening external events ($e$ types) for each instruction resulting in far fewer cases (O($m^2e^2$)) (e.g., a few thousands) than the number of the reorderings (e.g., $100!$), achieving  scalability in the number of in-flight memory instructions.
Thus, QED is  scalable in {\em both} the numbers of in-flight memory instructions and of cores for all MCMs. 

While the above contributions enable fewer instruction-event interleavings to be considered,  each such interleaving must be checked against the MCM (i.e., is the interleaving allowed?) and in the RTL implementation (i.e., does the interleaving occur?).
To that end, we propose a novel {\em observability-based} method. {\em Instead of checking whether the observed values from a reordered execution can also be produced by an MCM-permitted ordering, QED checks whether a reordered execution producing some observed values can be {\em restored}}
{\em to achieve an MCM-compliant ordering without changing the values.} That is, the restoration must not be {\em observable} by any instruction or external event.
For instance, the verifier cannot restore a load past an invalidation to the load address 
because such a restoration may change the load value 
(except for silent stores~\cite{silent-stores,incoherent}). A successful restoration means a 
given execution order and values are permitted by the MCM,
whereas a failure implies an MCM violation. 

\figput{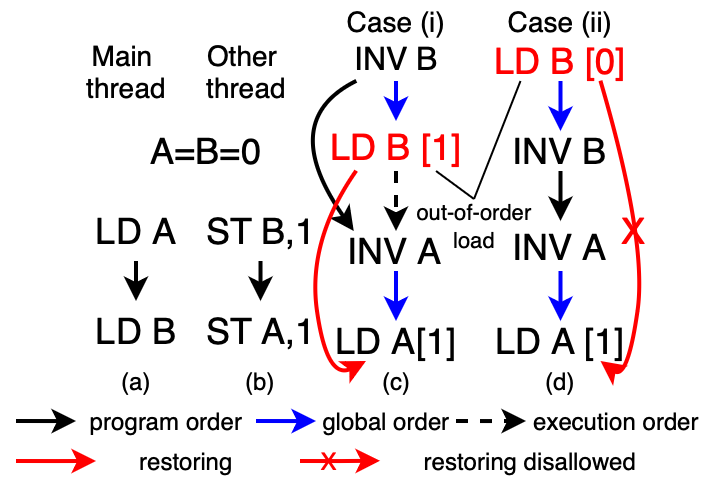}{}{SC example restoration}
 
Consider the simple SC example thread in~\figref{intro-eg}(a). 
Assuming {\em ld B} executes out of order before {\em ld A}, QED tests this execution by {\em introducing}  (external stores') invalidations ({\em inv}) to $A$ and $B$. Assume that the invalidations are from another thread, shown in~\figref{intro-eg}(b), which executes in program order. 
In the main thread's out-of-order execution, the external {\em inv B} (a proxy for {\em st B}) may be ordered globally (i) 
{\em before} {\em ld B}  (\figref{intro-eg}(c)),
or (ii) {\em after} {\em ld B} (\figref{intro-eg}(d)). 
To be brief,  we do not show other possible orderings of the external stores. Now, QED considers {\em only} the main thread's orderings  
{\em but not} how many other threads there are.   
In case (i) (\figref{intro-eg}(c)), because  {\em ld B}'s address {\em differs} from {\em inv A}'s and {\em ld A}'s, {\em ld B} can be  restored unobservably after  {\em ld A} as required by SC.
In case (ii) (\figref{intro-eg}(d)), because the top and bottom orderings affect {\em ld B} and {\em ld A} values, respectively, neither {\em ld B} nor {\em ld A} can be
restored unobservably, signaling an SC violation. 
(Squashing {\em ld B} only upon a later {\em inv B}, even without  {\em inv A},  is conservative and correct~\cite{two_techniques}.)

QED decomposes the problem of verifying an RTL implementation into two parts.
The first part
considers {\em all} possible pair-wise access reorderings  including an exhaustive set of intervening external events, 
which are few enough to remain tractable. Each re-ordered access pair and the intervening events represent an execution trace which QED tries to restore to conform to the MCM.
The exhaustive {\em exploration} is organized as a {\em tree} for each access pair where the traces result in  a {\em decision tree} of simple, narrowly-defined {\em predicates} (e.g., is a reordered load squashed upon an invalidation to the accessed block before the load commits?). The second part (future work) processes the RTL implementation to evaluate the predicates using some manual annotation to indicate the appropriate signals, automatic dataflow analysis of the implementation, and verification tools such as SMT solvers.

Thus, assuming the pipeline front-end and coherence are implemented correctly, QED scalably verifies the LSQ and coherence interface.
In our experiments, we automatically generate the exploration and decision trees for SC, TSO, and RISC-V WMO, and illustrate mechanical and automatable verification by evaluating a substantial predicate against BOOM v3 implementation of RISC-V WMO. We leave the full automation of the predicate evaluation to future work. We humbly point out that the ``*check'' series includes at least four papers~\cite{pipecheck,ccicheck,coatcheck,tricheck
} without RTL evaluation.
\putsec{background}{Background}

\putsubsec{modern_systems}{Modern systems}

\begin{figure}
    \centering
    \includegraphics[width=\linewidth]{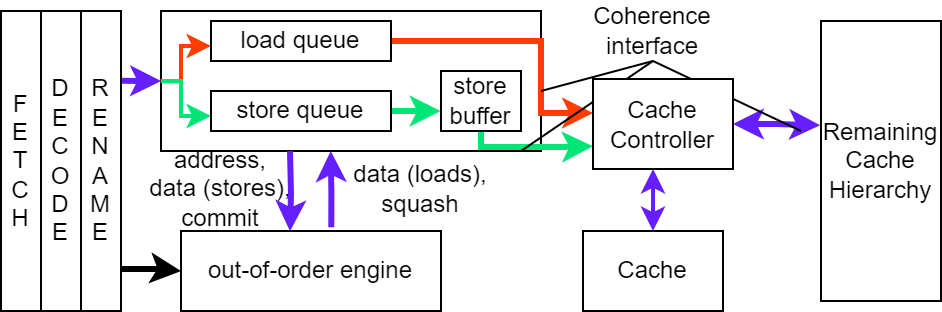}
    \caption{Data path of loads (red,blue) and stores (green,blue).}
    \label{fig:lsq}
\end{figure}

We consider a modern multicore system comprising out-of-order issue cores and multi-level memory hierarchy. While out-of-order issue processors have substantial mechanisms for speculation, register renaming, and out-of-order issue which focus on register dependencies, we focus  on memory instructions after they are issued.
As discussed in~\secref{intro}, we assume that bugs in the pipeline front-end related to register dependencies and global cache coherence (i.e., write serialization and, if required by the MCM, write atomicity) 
have been caught. We focus on memory ordering in the load-store queue and coherence interface where most consistency bugs occur (\figref{lsq}). 

\putsubsec{lsq}{Load-Store Queue and Coherence Interface}

Modern load-store queues (LSQs) in out-of-order issue processors reorder and overlap memory accesses (\figref{lsq}). While loads may be issued out-of-order to the cache,
a store is issued to the cache only when the store reaches commit to ensure precise interrupts (a store may prefetch coherence permissions  
before reaching commit). However, stores may be overlapped in the cache hierarchy in weaker MCMs and may complete out-of-order (e.g.,  store misses). 
A load  returns a value to the pipeline and is {\em globally ordered} after the
store that produced the value~\cite{dubois_correct}.
A store (a) is complete when the writer receives the acknowledgments of invalidations of all the copies and  (b) is performed locally to the cache. 
The ordering between these  two parts depends
on the consistency model (i.e., whether writes are atomic).
A store is {\em globally ordered} after (a) the store that produced the previous value and (b) the loads that read the previous value (well-defined because writes to {\em  one} location can be {\em serialized} in all MCMs~\cite{tutorial}). 

In addition to the LSQ, the coherence interface between global coherence and the node (comprising the core and local cache hierarchy) could also potentially violate the MCM (\figref{lsq}). 
The coherence interface (i) sends out requests for misses (including prefetches), (ii)
delivers external invalidations to the
LSQ and cache hierarchy, (iii) sends out local write invalidations, collects the acknowledgments, and sends the write to the cache, and
(iv) writes back dirty evicted blocks to lower levels of memory hierarchy.
The coherence interface must order outgoing (read and write) misses and incoming invalidations (e.g., on the local bus). 
QED verifies that the coherence interface preserves this local order.

\putsubsec{models}{A few common consistency models}

The most intuitive model is sequential consistency (SC)~\cite{lamport}. SC requires the  global memory order (\lm) to be a total order of all memory accesses to any location across all threads~\cite{synth-lecture}. We extend \lm to include external coherence events (e.g., incoming invalidations and external reads), which are proxies at a given thread for memory instructions in other threads.
Further, SC requires all accesses from a  thread in this global order to obey the thread's program order, denoted by \lp~\cite{synth-lecture} (solid black arrows in~\figref{intro-eg}). 
The global memory ordering implies that any load from  a location retrieves the value of the latest store to the location (``latest'' is well-defined as per the global order) (blue arrows in~\figref{intro-eg}). 
If there is no total \lm order or if the \lm order violates the \lp order (\figref{intro-eg}(d)), the system is not SC-compliant.

While SC is the simplest model (highest programmability), 
SC's 
strict ordering imposes performance overhead. Total Store Order (TSO) is a  commonly-used model which allows a load from a location to occur before previous 
stores to different locations in program order. Such a schedule helps hide
load latency and improves performance. A load  to the same location as a previous store must obey program order to enforce 
the store-to-load dependence. 
Load-to-load, load-to-store, and store-to-store program orders are not relaxed.

In more relaxed memory models, most program order constraints and, in some cases, write atomicity are relaxed~\cite{tutorial}. 
Instructions can be executed out-of-order if the addresses do not match -- data dependencies are still preserved.
Ordering among instructions and atomic writes are  programmed explicitly using some synchronization primitive -- atomic instructions, acquire/releases, or memory barriers.
The synchronization point denotes the time after which all threads are guaranteed to have seen all the instructions that executed since the last synchronization point.
Between two synchronization points, the memory model relaxes all constraints between loads and stores, maximizing performance.

Finally, some  MCMs may include address, data, or control-flow dependencies in ordering requirements (e.g., RVWMO).
For such MCMs, \name assumes that the  pipeline front-end handling register and control-flow dependencies is verified separately (discussed in~\secref{intro}), and that the front-end correctly marks such memory instructions in the LSQ so that \name can verify that the LSQ  meets the ordering requirements. 
\figputW[\linewidth]{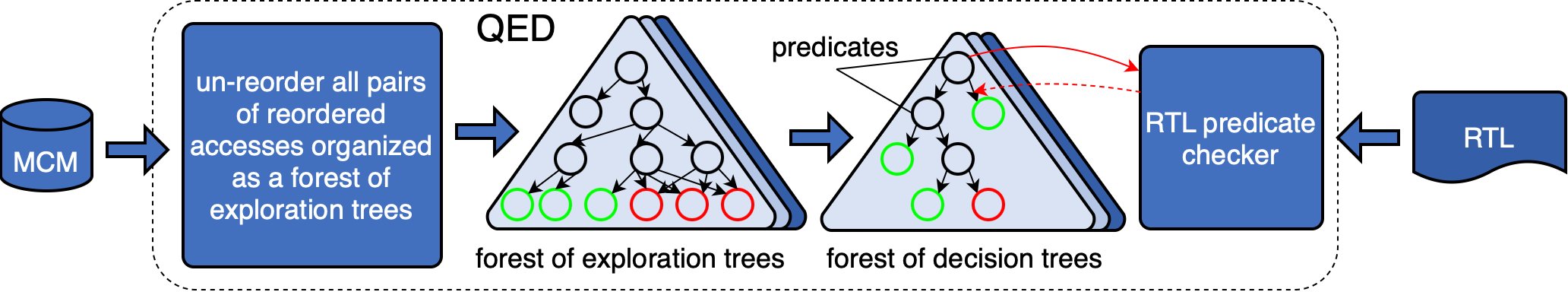}{}{QED}

\putsec{qed}{QED}

Recall from~\secref{intro} that we propose {\em QED}, a scalable method 
to verify an RTL implementation 
against a given MCM. QED focuses on the 
load-store queue (LSQ) and the coherence
interface between global coherence and the node (core and local cache hierarchy). 
Our key observations are: (1) Instead of considering all  reorderings of all in-flight memory instructions, only 
transitively non-redundant memory instruction pairs connected by an edge in the MCM-imposed partial order, called directly-ordered instruction pairs, 
need to be considered. (2) To consider the effects of other threads via external events at a given thread, we need to consider only the ordering of the events with the given thread's instructions and not where the events originate.    
These two observations allow QED to scale in both the numbers of in-flight memory instructions (i.e., the LSQ size or cache hierarchy parameters) and of cores. 
Specifically, we propose an {\em observability}-based approach which tests  
whether a reordered execution producing some observed values can be {\em restored} to achieve an MCM-permitted ordering  while remaining {\em unobservable} by any instruction or external event (i.e., without changing the values).  
To verify an RTL implementation, QED first exhaustively explores all possible pairwise instruction reorderings, including intervening external events, organized as a forest of {\em exploration trees} (\figref{overview}). QED then restores the reordered sequences giving rise to a {\em decison tree} of simple predicates which are evaluated by processing the RTL implementation. 

\putsubsec{pairwise}{Directly-ordered instruction pairs}
A memory consistency model (MCM) is defined by (1) the subsets of memory instructions among which program ordering must be preserved and (2) the appearance (or lack) of write atomicity~\cite{tutorial}. 
For example, sequential consistency requires the preservation of all program order and write atomicity.
Nominally, the model imposes ordering among  arbitrary number of memory instructions, which is the first scalability challenge for QED. 
Accordingly, QED's first key observation is that only the ordering between memory instructions connected by an edge in the MCM-imposed partial order
needs to be considered because it is impossible to violate the ordering of two arbitrary memory instructions in a thread {\em without also violating the ordering of two  memory instructions connected by an edge in the MCM-imposed  order}.

This observation is easy to show in SC because preserving ordering only between consecutive pairs ensures every other required order, due to transitivity. For example, consider three memory operations $a$ \lp $b$ \lp $c$. SC requires preserving all orders, denoted by $a$ \lm $b$, $b$ \lm $c$, and $a$ \lm $c$. However, preserving {\em only} the order between consecutive pairs ($a$ \lm $b$, and $b$ \lm $c$) is enough to guarantee all required order. 

To extend the above argument to MCMs that may  impose only partial ordering, we generalize as follows.
Two memory instructions $i$ and $j$, such that $i$ \lp $j$, whose ordering must be preserved are {\em directly-ordered} if the edge $(i,j)$ is in the {\em transitive reduction}~\cite{transitive-reduction} of the directed graph induced by the MCM's partial order. By definition, a transitive reduction eliminates all transitively-redundant orderings while  preserving all orderings
directly or indirectly (by transitivity), so that  a transitive reduction of a graph has the same transitive closure as the original graph. 

In relaxed models, 
the above extension captures all  directly-ordered pairs even though the memory instructions may have one or more intervening instructions. For example, 
consider instructions $a$, $b$, and $c$ consecutive in program order and the MCM-imposed orders 
$a$ \lm $c$ and $b$ \lm $c$ but {\em not} $a$ \lm $b$, then $a$ and $c$ are directly-ordered as well as $b$ and $c$.

Consider the following proof sketch for an MCM where two memory instructions $i$, and $j$ should be ordered  as $i$ \lm $j$ but are executed and observed by the memory system in inverted order (i.e., $j$ \lm $i$), which is a violation. 
In the graph induced by the MCM-imposed partial order, there is a directed edge from  $i$ to $j$ indicating that the order must be preserved. In the transitive reduction of the graph, the edge may be replaced with one (or more) path(s) from $i$ to $j$. Consider labeling all nodes
along one such path as $l_0, l_1, l_2,\ldots,l_n$ where $l_0$ is $i$ and $l_n$ is $j$. There are two cases to consider. In the first case of $n=1$, there is a direct edge between $i$ and $j$, so our claim is trivially proved because $i$ and $j$ are connected by an edge in the MCM-imposed partial order. 
In the second case, 
consider the first instruction $l_k$ where $0<k\le n$ along the path that violates the MCM with respect to instruction $i$ -- i.e., $l_k$ \lm $l_0$ (note, $l_k$  could be $l_n$).  $l_k$ must violate the required MCM  ordering with its preceding instruction on the path $l_{k-1}$ (which could be $l_0$) because either (1) $l_k$ \lm $l_0$ if $l_0$ is the same as $l_{k-1}$, or (2) if $l_0$ and $l_{k-1}$ are distinct, then $l_k$ \lm $l_0$ \lm $l_{k-1}$ because all orders from $l_0$ to $l_{k-1}$ before the first violation (at $l_k$) must be preserved, by definition. Therefore, we have proved 
that any arbitrary MCM-ordering violation {\em must} result in an
MCM  violation between a pair of memory instructions that are connected directly in the MCM-imposed  order ($l_{k-1}$ and $l_k$). 

Write atomicity, the second part of the MCM, is handled by the coherence interface, as discussed in \secref{civ}.

\putsubsec{scalability}{Scalability to any number of cores}

Any number of  cores executing arbitrary code may interact with a given core, which is the second scalability challenge for QED. 
However, these interactions occur in the form of external events at the chosen core (e.g., incoming invalidations and read requests). Fortunately, 
it is sufficient to consider only the ordering of instructions and external events independent of the events' originating cores because implementations do not consider the events' origins (e.g., an out-of-order load is squashed upon a matching invalidation  regardless of the invalidation's origin~\cite{two_techniques}).
Furthermore,  only pairs of directly-ordered memory instructions (based on~\secref{pairwise}) and intervening external events need to be considered.  
While the events' origins do not matter, the ordering among the events as well as those between the events and a given core's instructions do. 
Two external events (say two invalidations to addresses A and B, denoted as {\em inv A} and {\em inv B},  respectively) may be ordered (e.g., from the same thread) or  not ordered (e.g., from different threads or non-atomic stores) 
Further,  events from even different threads may be ordered  globally through a sequence of instructions on different cores. For instance, in~\figref{intro-eg}(d),  {\em inv B} \lm {\em LD A} where {\em inv B} is a proxy for {\em st B} in the other thread under our extended notion of \lm (\secref{models}). 
To be exhaustive, we consider all orderings of  external events (i.e., both {\em inv A} \lm {\em inv B} and {\em inv B} \lm {\em inv A}) as well as no ordering between the events.
Each of these cases means that {\em there exists} a valid code example under the
given MCM that produces the ordering. Further, because the events' origins do not matter, these cases capture all valid examples  that produce each ordering. 
Based on the pairwise observation in ~\secref{pairwise} and this discussion, QED is scalable in  both the numbers of in-flight memory instructions and  of cores. 

\putsubsec{obs}{Observability}

QED is based on the observation that an out-of-order instruction
violates a given MCM if and only if the out-of-order instruction is {\em observable -- i.e., the values produced by the out-of-order execution cannot be produced  by any  MCM-compliant execution}. 
From this definition, we can infer types of reordering that are not observable. For example, consider the lifetime of a new value that is loaded by a given thread/node: the lifetime begins when a new value is brought in due to a load miss (or prefetch), and ends when an invalidation to that address is applied (indicating that a potentially different value\footnote{Some optimizations have also been proposed for  silent stores where the same value is written~\cite{silent-stores,incoherent}} has been written elsewhere) (e.g.,  in~\figref{intro-eg}(d), {\em ld B}'s lifetime ends at {\em inv B}) or when there is a store within the thread. 
We can also conservatively consider that a value's lifetime ends on cache eviction because subsequent invalidations to that block are not delivered to the node. 
Within the lifetime of a value, reordering loads with respect to other instructions is not observable because the reordered load returns the exact same value. However,  no load can be reordered, without being observable,  
earlier than  the  store that produced the value, or 
later than the invalidation/store that  overwrites  the value. 

Analogously, a complete set of observability rules can be derived for loads and stores. A store is observable only by a load for the same address in the same or a different thread; and vice versa (e.g., in \figref{intro-eg}, {\em st B} in the other thread observes {\em ld B} in the main thread). 
Furthermore, reordering loads and stores across non-memory instructions is not observable for consistency purposes. (Of course, data dependencies must still be maintained for correct single-thread execution.)

\putsubsec{un-reorder}{Restoration}

QED uses the above notion of observability in its verification method as follows. 
For a given set of instructions in a  thread, consider the basic relationship between program order (\lp), the {\em executed} order (\lte) from an implementation (solid and dashed, black arrows in \figref{intro-eg}), and the required memory ordering  for an MCM-compliant system (i.e., the subset of valid \lm orders).
If the executed order \lo may be transformed {\em unobservably} -- {\em i.e., without changing the execution's values} --  into an MCM-compliant memory order \lm then 
the system is MCM compliant, even if the untransformed execution order seemingly violates the MCM.  

We refer to such unobservable transformations as {\em restoration} which undoes the effect of instruction reordering that an implementation may perform for power/performance  optimizations. Recall that \lm ordering inherently implies a global order  which affects some thread's values and therefore cannot be restored. Only \lo ordering may be  restored. Any out-of-order reordering uses  \lo (e.g., in~\figref{intro-eg}(c), {\em ld B} \lo {\em ld A}). Any external event can be ordered with any local memory instruction using \lo when there is such an execution order but no global ordering (\lm) is forced by the instructions and events. For instance,  assume~\figref{intro-eg}(a) and  the other thread is a singleton {\em st B} without any access to $A$ (unlike~\figref{intro-eg}(b)). Now, {\em st B}'s {\em inv B} may arrive at the main thread so that {\em ld B}  \lm  {\em inv B} \lo {\em ld A} which can be restored.  This scenario is different from~\figref{intro-eg}(d), where  {\em inv B} \lm {\em ld A} is forced by the other thread in~\figref{intro-eg}(b).

In contrast, \lm ordering occurs in the following cases. 
(1 -- one thread): Any \lp ordering, involving same or different addresses, required by the MCM and obeyed by the execution  is an  
\lm ordering by default. (2 -- one address): Any load and invalidation, invalidation and invalidation, store and external read pairs to the same address can be ordered (or serialized) by \lm to  imply a particular ordering (e.g.,  in~\figref{intro-eg}(d), {\em ld B} \lm {\em inv B}). 
Any restoration past such ordering
may change execution values (e.g., 
a load cannot be restored past an invalidation to the load address because 
the load value may change). Combining these two cases for multiple threads and different addresses, \lm 
ordering at one thread involving different addresses  and at least one external event captures MCM-relevant \lp ordering involving  those addresses in another thread which is invisible to the first thread.
For example, in~\figref{intro-eg}(d), 
{\em inv B} \lm {\em ld A} is implied by~\figref{intro-eg}(b), where {\em st B} \lp {\em st A} (in the other thread) \lm {\em ld A} (in the main thread). 
To be exhaustive, QED tries all valid combinations of \lm orderings in each case (e.g., in~\figref{intro-eg}(c) and (d), {inv B} \lm {\em ld B} and {\em ld B} \lm {\em inv B}, respectively).  
Some combinations may be impossible, as explained in detail in~\secref{proof-frame}.    

While~\figref{intro-eg} shows an SC example,  MCMs with non-atomic  writes or relaxed write-to-write order do not impose {\em st B} \lp {\em st A}, leaving only {\em inv B} \lte {\em inv A} to consider for such writes. 
However,  such MCMs do impose order via fences (within a thread) or atomic writes (across threads), so that a valid,  {\em adversarial} case where {\em st B} \lp {\em st A} (e.g., due to a fence) exists(~\secref{scalability}).
\name proves correctness of a hardware implementation, not of a specific test example. Therefore,  \name must consider all cases.

In addition to \lm orderings, we exhaustively consider  all \lte orderings. However,  if a \lm ordering does not produce a violation then \lte cannot because \lte orderings can be restored unlike \lm orderings. Hence, a \lte ordering between a pair of external events is subsumed by a \lm ordering between the same events so that only the latter needs to be considered. 

The target of QED's restoration are MCM-compliant orders. 
Such MCM-compliant \lm order(s) must preserve some partial order from the \lp program order (with SC being an exception that requires valid \lm orders to preserve all the \lp orders). 
In general, a valid MCM-compliant order is any topological sort of the graph induced by the MCM's partial order.

QED generates all possible execution orders (\lte) of pairwise instructions exhaustively combined with external events (e.g., incoming invalidations and external reads, and outgoing misses). 
Because the number of external event types and memory instruction types are  small (e.g., $<10$), the number
of orderings remains tractable (complexity  analysis in~\secref{auto}). 

For each such execution order, QED restores instructions which execute out-of-order (i.e., restores \lte  order) to bring back an MCM-compliant \lm  order without being observable. 
Adhering to the restoration rules, if no restoration can arrive at a valid MCM \lm order without being observed, the execution  violates the  MCM.
    
\putsubsubsec{simple-egs}{A sample restoration for SC}

\figput{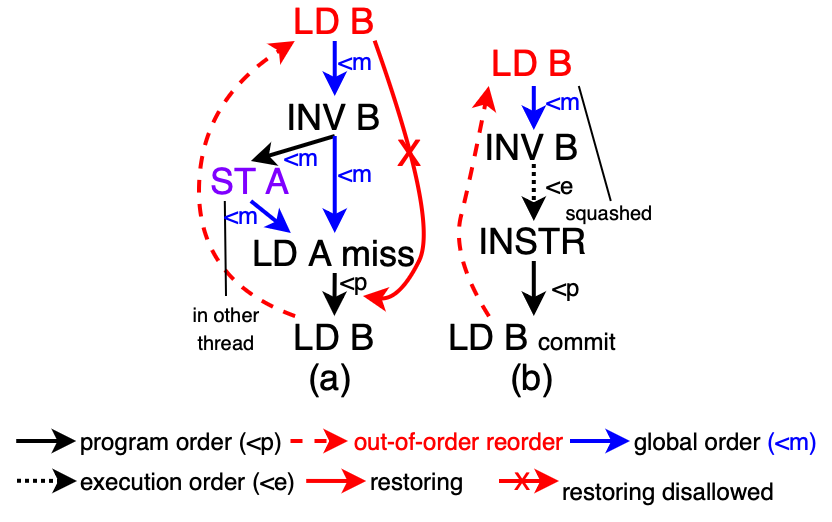}{}{Example restoration}

We consider out-of-order loads in SC implementations that have mechanisms to discover and squash unsafely-ordered instructions.
Recall that in~\figref{intro-eg}(d), {\em ld B} \lm {\em inv B} \lm {\em inv A} \lm {\em ld A}, which cannot be restored.
Instead of {\em inv A}, we consider {\em ld A miss} which may bind  a  new value for $A$ from another thread's {\em  st A} where {\em  st B} \lp {\em  st A} which is invisible to the first thread (\secref{un-reorder}). See \figref{unreordering-egs}(a). Then, in the first thread we have  {\em ld B} \lm {\em inv B} \lm {\em ld A miss} which also cannot be restored.
To cause a violation, both cases ({\em inv A} or  {\em ld A miss})  require {\em at least} an {\em invalidation (inv) B} to the thread after {\em ld B}  but before {\em ld B commit}, (i.e., {\em ld B} \lm {\em inv B} \lo {\em ld A} \lp {\em ld B commit}). 
However, this execution sequence would be squashed. Only the 
execution sequences that do not contain an {\em inv B} between load issue and commit would commit, i.e., {\em ld B} \lo {\em ld A} \lo {\em ld B commit}.
In such  sequences, however, the out-of-order {\em ld B} is not observable and can be restored next to  {\em ld B commit}. 
Note that while
{\em inv A} (or {\em ld A miss}), and  {\em inv B} are needed to show an SC violation, 
squashing upon {\em inv B} alone (\figref{unreordering-egs}(b)), to simplify the implementation, is enough to prevent the violation~\cite{two_techniques}.

\putsubsec{proof-frame}{Verification framework (Illustration with SC)}

QED takes two inputs - the MCM specification and RTL implementation to be verified. The MCM specification provides constraints on relaxations of orderings and atomicity with which the memory instructions  have to comply in any valid execution.
For example, in SC, the MCM requires that {\em ld-ld, ld-st, st-ld} and {\em st-st} program order is captured in the execution order and there exists a  global order among them.

Our proof method generates an exhaustive, hierarchical, list of {\em execution traces} for each pair of instruction type (with same or different addresses),  organized as a tree. 
This list includes relaxations in orderings, exploring cache hits, misses and store-load forwarding for each instruction and enumerating every external event, which can potentially result in a violation. 
Though an instruction may have many observer event {\em types} (e.g., invalidations, misses, prefetches, and evictions for loads), considering 
multiple event types for the same instruction is redundant because the
all the event types result in the same observation about the instruction.
Thus, we need to consider only one such event type per instruction at a time though
all the event types do have to be considered (but not together).
Further, we need to consider only a single event of any type. If a memory instruction can(not) can be restored past an event, then the instruction can(not) can be restored past multiple events of the same type.
While \secref{pairwise} shows that only  directly-ordered pairs of instructions need to be considered, out-of-order reordering may move other instructions in between such a pair. However, we consider only external events but not these other instructions 
because  these instructions are in the set of all types of instruction pairs we consider. Thus, these other instructions are covered as well. 

\figput{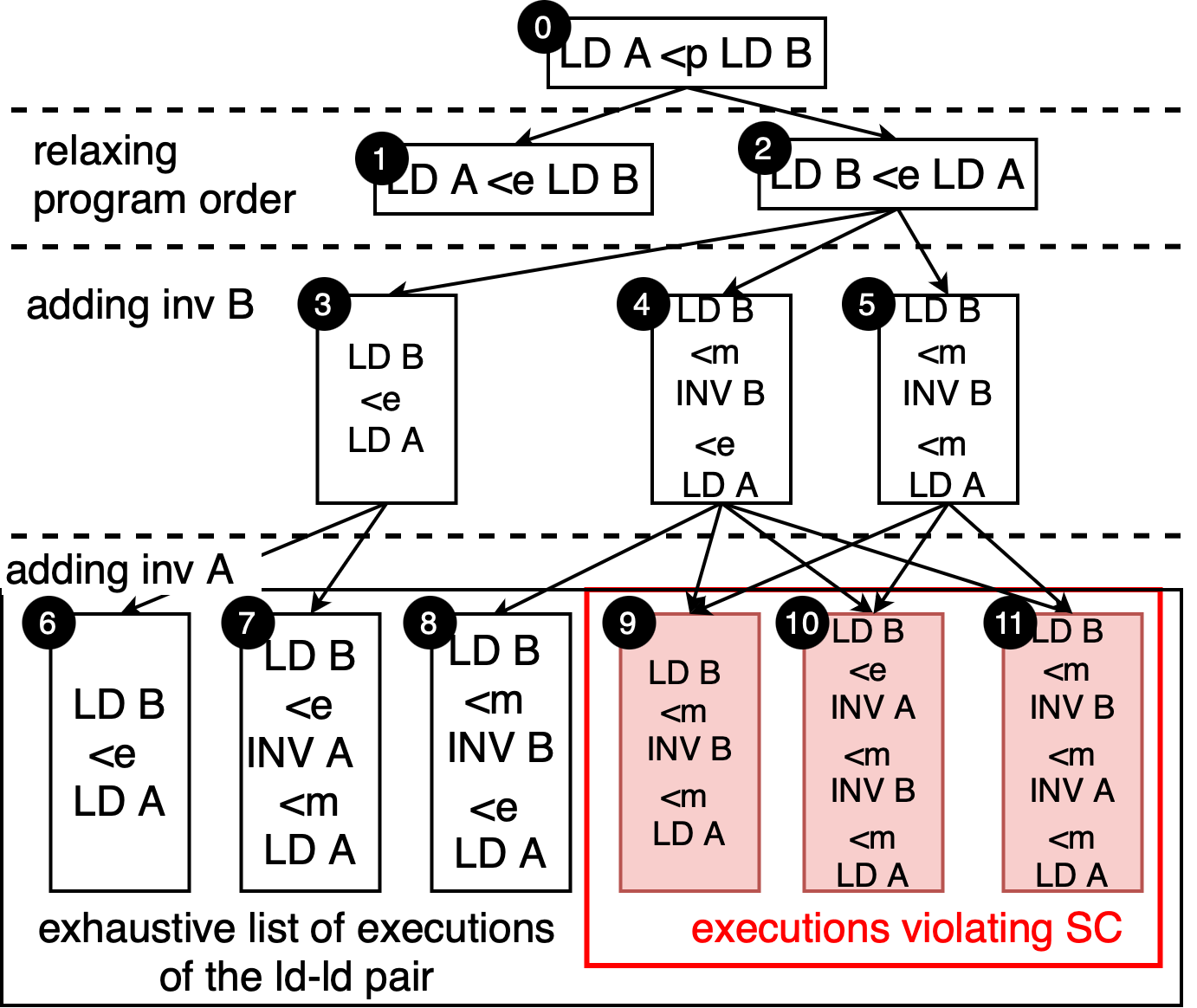}{}{Exploration tree for ld-ld ordering in SC showing only invalidations}

\figref{ld-ld-tree} shows such an exploration tree for the {\em ld-ld} pair with different addresses. In the figure, we progressively and exhaustively add {\em inv B} and {\em inv A} and their relative orderings. Recall from~\secref{un-reorder} that \lte is subsumed by \lm and hence not considered. {\em Node 2} expands into {\em nodes 4} and {\em 5} by adding {\em inv B} with \lte and \lm orderings with {\em ld A}, respectively,  as the {\em ld}  and {\em inv} are to different addresses ({\em ld B} and  {inv B}, to the same address, have only \lm choice).  While \lte ordering is always possible,  not all \lm orderings are possible even for different addresses. For instance, in {\em node 7} {\em inv A} is added to give {\em ld B} \lte {\em inv A} \lm {ld A} without expanding  {\em ld B} \lm {\em inv A}  \lm {ld A} as a node.
Being to different addresses,
{\em ld B} \lm  {\em inv A} requires  an intervening  {\em inv B}  such that  {\em ld B} \lm {\em inv B} \lm {\em inv A},  corresponding to, say,  {\em external st B} \lp {\em external st A} in another thread, in the absence of an eviction of $B$ from the main thread's cache between {\em ld B} and {\em external st B}. 
However, {\em node 7} does not include an  {\em inv B}, resulting in only  {\em ld B} \lte {\em inv A}. 
In {\em node 5}, in contrast,  {\em inv B} \lm {\em ld A} is possible assuming {\em external st B} \lp {\em external st A} in another thread so that {\em ld A} misses and fetches {\em external st A}'s value even without any intervening {\em inv A} (as discussed in~\secref{simple-egs}).  
{\em Node 11} shows the same case with an intervening {\em inv A}. 
Note that in {\em node 7}, assuming {\em ld B} miss would not change anything. {\em Node 10} is a subtle case that adds {\em inv A}  using a \lte ordering with {\em ld B} for the same  reason as {\em node 7} (which needs {\em inv B} \lm {\em inv A} for the \lte ordering to become \lm but {\em node 10} has {\em inv A} \lm {\em inv B}). Further,  {\em node 10} inherits {\em inv B} \lm {\em ld A} from {\em node 5} assuming a {\em ld A} miss which fetches the value from an {\em external st A} which is ordered {\em after} {\em inv B} (similar to {\em node 11}). However, {\em inv A} is ordered {\em before} {\em inv B} and hence is from a {\em different} external store. We consider prefetch and eviction in~\secref{prefetch}.

At the leaves, we apply the restoration rules to each execution trace to achieve a valid MCM \lm order.
Each trace that cannot be restored violates the MCM.
\figref{ld-ld-tree}  shows three such  execution traces, highlighted in red. While the violations are straightforward (\lm cannot be restored), {\em node 8} is not a violation because {\em inv B} \lte {\em ld A} (different addresses) can be restored so that {\em ld A} moves past {\em inv B} and again past {\em ld B} to produce the  \lp order.  In {\em node 10}, {\em ld B} can be restored past {\em inv A} which
is not relevant to, and cannot prevent, the violation. 

\putsubsubsec{prefetch}{Prefetch and eviction}

Coherence prefetch is usually thought to be safe because stale prefetched values are invalidated when new writes occur. However, unrestricted  prefetches  can  make load reordering incorrect. For example,  
in~\figref{unreordering-egs}(a), 
a prefetch may prevent a miss which otherwise would've led to a squash (e.g., instead of ld A miss in~\figref{unreordering-egs}(a), ld B \lm inv B \lm prefetch A \lm ld A hit ).
Fortunately, treating  prefetches as misses (to 
 order after the store to the same address which begins the value's lifetime) 
cleanly handles these issues. Evictions can be treated similar to  invalidations (as the end of a value's lifetime). 

\putsubsubsec{full}{Atomic and non-atomic MCMs}
\figput{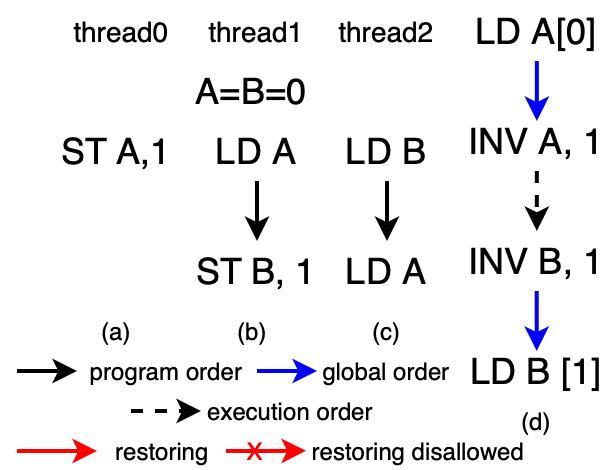}{}{Causality test involving stores. Assume the instructions in {\em thread1} and {\em thread2} are ordered within each thread as per the MCM or using some fences. (d) shows out-of-order execution of {\em thread2} in PC where the invalidations are not ordered, allowing restoration.}

While our examples show restoration in one thread, 
some consistency tests involve multiple threads where both write atomicity and program order matter (e.g., the causality test in~\figref{causal} where two writes from different threads are related causally and are read by a third thread). In such cases, write serialization and, if required by the MCM, write atomicity (e.g., {\em thread0} in \figref{causal}(a)) are covered separately by standard coherence verification  (this separation is discussed in~\secref{intro}); and program order in each thread is checked by one of \name's exploration trees. In~\figref{causal}(b) and~\figref{causal}(c), {\em thread1} and {\em thread2} are covered by  {\em st-ld} and {\em ld-ld} trees, respectively. 

Also, \name seamlessly handles non multi-copy-atomic MCMs (e.g., processor consistency (PC)). Such MCMs, where  writes cannot be ordered globally, do not impose \lm ordering on the writes, as discussed in~\secref{un-reorder}. For example, assuming PC in~\figref{causal}(d), the invalidations to {\em thread2} are ordered by \lte  and not \lm. Thus, {\em ld A} \lm {\em inv A} can be restored past {\em inv B} \lm {\em ld B} to achieve the PC-compliant ordering of {\em ld B} \lm {\em ld A}.
In such MCMs, however, atomic writes or ordered non-atomic writes in one thread (ordered either by the MCM rules or via fences) do obey \lm ordering.
Therefore, \name can impose \lm ordering 
even in these MCMs, because such valid, adversarial cases are possible (e.g., if both {\em st A} and {\em st B} are  atomic in~\figref{causal}).  \lm ordering is absent only in an (impractical) MCM that has neither atomic writes nor fence-like ordering.

\putsubsubsec{auto}{Automating the framework}

We  automate the generation of the exploration trees by considering  all pairwise memory instruction types (which results in a tree for each such pair) while also considering all possible interleavings of external event types (which adds nodes to the trees). Algorithm~\ref{alg:treegen} generates out-of-order execution traces for each instruction pair (each consistency rule considers a pair). The algorithm enumerates the interleavings of {\em observer events}, relevant to the given pair of instructions (e.g., 
invalidations  and external reads observe loads and stores to the same address, respectively, as well as cache controller events such as misses, prefetches, and evictions).
{\tt enumerate(trace, event)} generates all permutations containing the instruction pair and events from {\tt trace} and {\tt event}.

\RestyleAlgo{ruled}
\SetKwInput{kwInput}{Input}    
\SetKwInput{kwOutput}{Output}
\SetKwComment{Comment}{/* }{ */}
\SetKwIF{If}{ElseIf}{Else}{if}{}{else if}{else}{}
\SetKwFor{For}{for}{do}{}
\DecMargin{1em}
\begin{algorithm}
\caption{\label {alg:treegen} Generating out-of-order execution traces}
\kwInput{MCM rule: A \lp B $\implies$ A \lm B}
\kwOutput{$traces$ = List$<$Tree$<$Traces$>>$}
$traces \gets \{\}$\;
\For{$oeB \in ObserverEvents(B)$}{
    \For{$oeA \in ObserverEvents(A)$}{
        $tree \gets Tree()$\;
        \If{B \lte A is allowed}{
            $tree.root(B $\lte$ A)$ \Comment*[r]{root}
            \For{$node \in leaves(tree)$}{
                $tree.add(node, duplicate(node))$\;
                \For{$seq \in enumerate(node, oeB)$}{
                    tree.add(node, seq)
                }
            }
            \For{$node \in leaves(tree)$}{
                $tree.add(node, duplicate(node))$\;               \For{$seq \in enumerate(node, oeA)$}{
                    tree.add(node, seq)
                }
            }
        }
        $traces.add(tree)$
    }
}
\end{algorithm}
\IncMargin{1em}

Assuming $m$ memory instruction types, there are at most $2m^2$ pairs (same and different addresses).
For each instruction type $i$, $e_i = size(observer-event(i))$ denotes the number event types (e.g., misses and evictions) that can observe $i$.
$e$ denotes $max_i(e_i)$.  
Of these $e$ event types, each trace has only one event type per instruction type (~\secref{proof-frame}).
Therefore, each instruction pair generates $O(e^2)$ trees, amounting to $O(m^2e^2)$ trees for the entire model (e is well under 10). 
This number is independent of the numbers of in-flight memory instructions and of cores.

At each tree leaf, we  automatically apply the restoration rules to the associated trace, but also manually check the results. The trace is a linear list of an out-of-order
instruction followed by the relevant external events and  another instruction.  Such automatic restoration (Algorithm~\ref{alg:unreorder}) simply moves  the first (last) instruction down (up) as far as possible freely past any \lte ordering  but not \lm ordering.
However, the instruction and any \lm-ordered event chain can be moved together while preserving the \lm order. 
The algorithm terminates  when either the first instruction (and its chain) moves past the last instruction and no violation is flagged, or 
the first instruction ends up in the same  chain as the last instruction and a violation is flagged because the instructions cannot be restored to an MCM-compliant order. 

Each trace has 4 items (2 instructions and 2 events) leading to $4!$ leaves per tree ($O(m^2e^2)$ trees total). The restoration of each leaf is nominally linear in the number of items 
because the first (last) instruction moves past or fuses with an event in every iteration, until the algorithm terminates. However, the algorithm restores only 4 items. Because of these constants, QED's overall complexity for the first step of exhaustive exploration is  $O(m^2e^2)$.

\RestyleAlgo{ruled}
\SetKwInput{kwInput}{Input}    
\SetKwInput{kwOutput}{Output}
\SetKwComment{Comment}{/* }{ */}

\DecMargin{1em}
\begin{algorithm}
\caption{\label{alg:unreorder} Restoring execution trace to \lp order}
\kwInput{Execution trace T: B $<_x$ $e_1 \ldots e_n$ $<_x$ A}
\kwOutput{valid: Bool}
\SetKwIF{If}{ElseIf}{Else}{if}{}{else if}{else}{end if}
\SetKwRepeat{Do}{do}{while}
\Do{B \lte A}{
B$^+$ = longest(B[\lm $e_i$ \lm $e_j$ \lm $\ldots$ \lm $e_k$]) $\in$ T\;
A$^+$ = longest([$e_p$ \lm $e_q$ \lm $\ldots$ \lm $e_r$ \lm]A) $\in$ T\;
\lIf{B$^+$==A$^+$}{valid=false; return}
\lElseIf{($\exists$ B$^+$ \lte $e$)}{
    T.rewrite(B$^+$\lte $e$ $\rightarrow$ $e$ \lte B$^+$)
}
\lElseIf{($\exists e$ \lte A$^+$)}{
    T.rewrite($e$ \lte A$^+$ $\rightarrow$ A$^+$ \lte $e$)
}
}
valid=true;

\end{algorithm}
\IncMargin{1em}

\putsubsec{RTL}{Predicate evaluation of RTL}

Based on the exploration tree and the execution traces that cause a violation (\figref{ld-ld-tree}), we generate a decision tree of a set of predicates (binary-response questions).
Each predicate infers if a certain relaxation or safety check is implemented in the microarchitecture, and is generated iteratively based on the answers to the previous predicates.
\figref{ld-ld-predicate} contains the predicates for {\em ld-ld} pair for SC and the order in which they are posed.
The first predicate asks if the processor implementation ever reorders loads to different addresses. If not, there can clearly be no {\em ld-ld} ordering violation. In implementations where {\em ld-ld} reordering is possible, the next predicate asks if such an out-of-order load ({\em ld B} in \figref{ld-ld-predicate}) is squashed upon receiving an invalidation to $B$~\cite{two_techniques}. 
Finally, even in designs where {\em ld B} is not squashed, it is possible to avoid a consistency violation if {\em ld A} is a hit which then can be restored above {\em ld B}.
But if {\em ld A} is a miss, then a  violation is unavoidable. 
Each such claim in response to a predicate (e.g., that no load reordering is allowed in the implementation, or that the {\em inv B} triggers a squash of {\em ld B}) 
is verified against the RTL. 

\figput[0.8\linewidth]{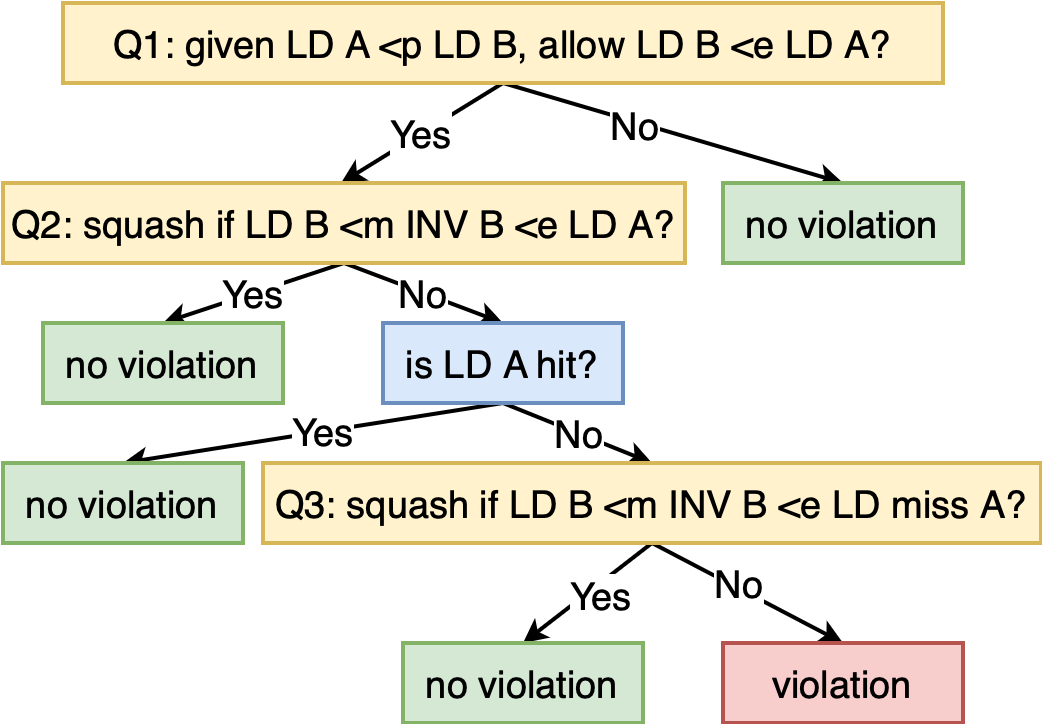}{}{Decision tree of predicates for load-load pair only with invalidations}

Once the decision tree of predicates is constructed, the target RTL implementation can be verified by checking if each predicate is satisfied and if the combination leads to a leaf of ``violation''. 
We envision a mechanical and automatable process which makes use of architect-provided metadata.

To verify the second predicate ($Q2$) in~\figref{ld-ld-predicate}, for instance, 
the architect is expected to map the relevant signals in the LSQ and the cache controller to the corresponding variables/fields of the RTL code. Note that correctness of the underlying building blocks is not within QED's scope. That is, we assume that a Content Addressable Memory (CAM) search circuit correctly searches the CAM. For $Q2$, a correct MCM implementation may conservatively employ  an LSQ search to detect a violating instruction, and a squash of the instruction and its dependent instructions if one is found. QED's predicate checking is limited to checking that the appropriate CAM search is presented to the LSQ and that a squash signal for the relevant instruction is raised if the search yields a match.  
A less-conservative implementation (whose correctness depends on $Q3$ in~\figref{ld-ld-predicate}) may squash only upon an LSQ match {\em and} a LD A miss. In this case, QED's predicate checking would confirm that a squash is triggered upon a match in an appropriate LSQ search and an LSQ-recorded miss for a relevant instruction. With the user-provided labels, every signal can be mapped to a corresponding RTL variable (e.g., a per-instruction field in the LSQ). 
Then the predicate checking can be done mechanically using automatic verification techniques (e.g., predicate abstraction-based model checking~\cite{slam,blast}).
See \secref{predicate-results} for the manual steps of an example. 
Moreover, because the amount of potential state in the RTL implementation is manageable (e.g., 100-300 entries in the LSQ), 
the verification is expected to be scalable with an encoding to SMT (Satisfiability Modulo Theories) solving. 

An implementation is verified to be correct if no violation is found in either restoration or predicate-evaluation step.
A violation in the restoration step denotes a high-level design bug
(e.g., missing a squash upon a certain invalidation).
A violation in the predicate-evaluation step 
is a low-level implementation bug
(e.g., squash not flushing the LSQ).

More advanced optimizations may buffer  coherence messages, reorder their application, and ensure correctness possibly via knowledge of global coherence ordering 
(e.g., Weefence~\cite{weefence}). QED's RTL checking can cover such optimizations by expanding the predicates.

\putsubsec{civ}{Coherence interface verification}
To ensure that the coherence interface does not introduce consistency bugs  by reordering events (even though the coherence protocol may be correct), we  verify that the coherence interface (1) preserves the local bus order (e.g.,  the actual application of the invalidation must be  ordered locally but the invalidation acknowledgment can be sent out of order) and (2)  orders  collecting the invalidation acknowledgments, and writing to the cache hierarchy and allowing other threads' early reads of partially-invalidated writes,  as per the MCM's write atomicity constraints.
Currently, because simple rules govern event ordering, verifying the interface using QED is straightforward. 
For future optimizations that  reorder events  at the coherence interface, we can extend our methodology to restore them unobservably.
\putsec{method}{Evaluation Methodology}

QED has two components: the MCM-based exploration trees leading to the decision trees of predicates and predicate evaluation against the RTL implementation.

\tabput{mcm}{
\begin{tabular}{|p{1.2cm}|p{2.6cm}|p{3.5cm}|} \hline
    MCM & Instructions & Ordering  \\ \hline
    SC & load, store, and synch. & all pairs order and synch. similar to store\\ \hline
    TSO & load, store, and synch. & relax store-to-load  order except synch. and store-to-load bypass relaxes write atomicity \\ \hline
    RVWMO & load, store, atomic, load-reserved, and store-conditional & relax all  order except  fences and acquire-release annotations \\ \hline
\end{tabular}
}{MCMs studied}

For the first component, we automatically perform QED's exhaustive exploration and automatically generate the trees for SC, TSO, and  RISC-V's WMO models (\tabref{mcm}). In SC, we consider loads, stores, and synchronization primitives which are identical to stores for ordering purposes.
We also consider read-own-writes early (store-load forwarding) which is legal in SC under  certain cases (e.g., absence of intervening misses). 
Compared to SC, TSO relaxes only the store-load program order for different addresses where store-load forwarding may lead to non-atomic writes to different addresses.  Atomic operations disallow
any program-order relaxation. In RISC-V's WMO consistency model, the configurable {\em fence} instructions  allow various ordering behaviors based on 4-bit annotations which order prior/later load/store instructions with respect to each other. Note that loads and stores are {\em not} ordered with respect to the fences themselves; fences {\em indirectly} order loads and stores with respect to each other. Further, because there is no ordering among the fences themselves, QED needs to consider ordering only between loads and stores.
Atomic operations, load-reserved, and store-conditional instructions support acquire-release annotations, similar to $RC_{SC}$~\cite{RC-SC}. These ordering constraints increase the number of exploration and decision trees to several tens which remains easily tractable. 

While we have automated the first component, as discussed in \secref{RTL}, the second component is automatable but not implemented in this paper. We consider RISC-V's WMO implemented in out-of-order-issue BOOM v3 (18K lines of Scala, 500K lines of generated Verilog) for demonstration because we did not find suitable out-of-order-issue implementations for SC and TSO. 
Fortunately, BOOM v3 is 
publicly available~\cite{boomprocessor} and is 
implemented in easy-to-understand, high-level and compact Chisel~\cite{chisel} making manual demonstration feasible. Even so, because manually verifying all the
relevant predicates is infeasible, we manually verify one substantial predicate -- load-load ordering for the same address which RISC-V WMO requires to be in program order even without any intervening fences or synchronization operations. BOOM v3 issues such loads out of order and squashes the later load upon an intervening invalidation to the load address before commit.
We include BOOM v3's relevant module, signal and code details in the results for ease of reading and 
a discussion on the steps needed to automate the process.
\putsec{results}{Results}
We present results for QED's two parts: exploration and decision trees and predicate evaluation of the RTL implementation. 

\putsubsec{tree-results}{Exploration and decision trees}

\tabput{dtreecount}{
\begin{stripetabular}{cccc}
MCM & \# Trees & \# Trivial trees & \# Leaves \\
\hline
\hline
SC & 24 & 5 & 103\\
TSO & 34 & 15 & 113\\
RISC-V & 112 & 69 & 305
\end{stripetabular}
\vspace{-0.2in}
}{Exploration/Decision tree counts for MCMs}

\tabref{dtreecount} shows the number of trees and leaves for the various MCMs. 
For SC, enumerating all pairs of loads and stores for same or different addresses, along with all external events, gives us 24 trees, some of which are trivial
(e.g., for {\em ld A} \lp {\em st B}, the store executes in-order after the load commits due to precise interrupts).
We distinguish between
same and different addresses for two reasons.
First, weaker models relax ordering among accesses to different addresses but require ordering when addresses match.
Second, even if the MCM ordering rules 
do not differ based on addresses,
implementations may treat the instructions differently.
The separate trees 
capture optimizations 
specific to each consistency rule.
Similarly, three types of instructions in TSO  (loads, stores, atomics) and five in RISC-V WMO (loads, stores, atomics, load-reserved and store-conditional), after enumerating external events, lead to 34 and 112 (including fence-based ordering) trees, respectively.

\vspace{-0.1in}

\putsubsec{predicate-results}{Predicate checking demonstration}
\figput[0.8\linewidth]{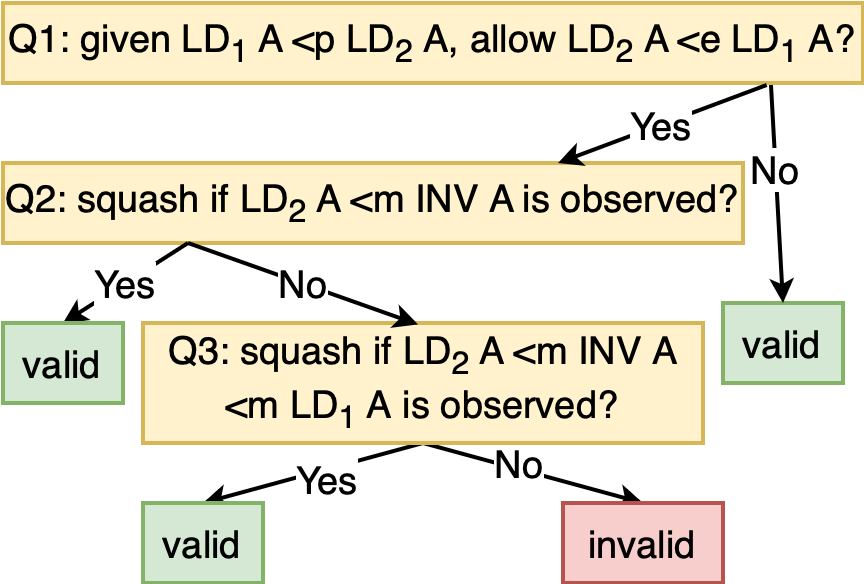}{}{Decision tree for RISC-V WMO ld A-ld A pair}

Below we demonstrate the manual steps we follow for checking a substantial predicate in BOOM v3: {\em ld$_1$ A} \lp {\em ld$_2$ A} $\implies$ {\em ld$_1$ A} \lm {\em ld$_2$ A}, where the subscripts distinguish the two loads to the same address $A$.
This procedure can be automated in the future for checking all predicates.
As discussed above, {\em ld$_2$ A} \lte {\em ld$_1$ A} is possible in BOOM v3 which squashes {\em ld$_2$ A} if {\em ld$_2$ A} \lm {inv A} \lm {\em ld$_1$ A}. 
Based on QED's exploration for the ld A-ld A pair similar to~\figref{ld-ld-tree}, \figref{rvwmo-proof-tree} shows the corresponding decision tree. The key predicate is: if {\em ld$_1$ A} \lp {\em ld$_2$ A} $\land$ {\em ld$_2$ A} \lm {inv A} \lm {\em ld$_1$ A} then squash {\em ld$_2$ A}.

First, we map each atomic condition relevant to the predicates to the corresponding variables/fields of the implementation. 
Specifically, {\em ld$_1$ A} \lp {\em ld$_2$ A} is captured implicitly by the load queue,  {\tt ldQ}, which holds and searches instructions in program order. 
The load address, load execution, and invalidation to a load address are captured in BOOM v3 RTL, respectively, by the signals/variables {\tt addr}, boolean {\tt executed} and boolean {\tt observed} fields per instruction in the {\tt ldQ}. Invalidations from the coherence interface are received in {\tt io.release} and the global signal {\tt ld\_xcpt\_valid} triggers a  squash.  Thus, 
{\em ld$_2$ A} \lo {\em ld$_1$ A} $\equiv$  {\tt ld$_1$.executed} $==$ $0$ $\land$ {\tt ld$_2$.executed} $==$ $1$.
{\em ld$_2$ A} \lm {\em inv A} $\equiv$ {\tt ld$_2$.addr} $==$ {\tt io.release.addr} $\land$ {\tt ld$_2$.executed} $==$ $1$ so that  {\em inv A} searches the {\tt ldQ} with $A$ and sets matching {\tt ld$_2$.observed} to $1$. Recall from~\secref{RTL} that we assume that circuit-level {\tt ldQ} functionality, such as
indexing, searching, writing and reading, are correct. 
Once the relevant variables/fields are identified, we convert each predicate to a ``Reachability Goal'' as illustrated in \tabref{rvwmo-ld-table}. Now our verification task is reduced to checking whether any execution can satisfy these reachability goals (in which all variables are implicitly existentially quantified).

Second, we manually verified that {\tt Goal$_3$} (which corresponds to the red "invalid" leaf 
in \figref{rvwmo-proof-tree}) is unreachable. Specifically, each of the predicates along the decision tree can be verified, because the RTL guarantees that 
(1) out-of-order execution is possible ({\tt Goal$_1$}), (2) invalidations are received correctly and noted in the LSQ state ({\tt Goal$_2$}), and (3) out-of-order loads (e.g., {\tt ld$_2$} in the above example) are always squashed once an invalidation has been received ({\tt Goal$_3$}). Note that 
{\tt Goal$_3$} is stated in an inverted fashion. The goal requires successful execution of the first load (without squash)  after meeting the pre-requisite goals,  an unreachable state. 

\tabput{rvwmo-ld-table}{
\begin{tabular}{|p{2.3cm}|p{5.5cm}|} \hline
    Predicate & Reachability Goal \\ \hline
    {\em ld$_2$ A} \lo {\em ld$_1$ A} is allowed?
    &
    {\tt Goal$_1$: i < j $\land$ ldQ[i] = A $\land$ ldQ[j] = A $\land$ ldQ[j].executed $\land \neg$ ldQ[i].executed } \\ \hline

    does not squash {\em ld$_2$ A} \lm {\em inv A}? 
    & {\tt Goal$_2$: Goal$_1$(i,j,A) $\land$ io.release.addr = A $\land$ ldQ[j].observed $\land$ $\neg$ld\_xcpt\_valid} \\ \hline

    squashes {\em ld$_2$ A} \lm {\em inv A} \lm {\em ld$_1$ A}?
    & {\tt Goal$_3$: Goal$_2$(i,j,A) $\land$ ldQ[i].executed $\land$ $\neg$ld\_xcpt\_valid}  \\ \hline
\end{tabular}
\vspace{-0.2in}
}{RISC-V WMO ld A - ld A predicate evaluation}

\putsubsec{discuss}{Future automation} 

We envision automating the predicate checking in the future. 
For the variable-mapping step,
we will require the architect to annotate the RTL implementation by identifying the RTL variables relevant to the predicates (e.g., {\tt addr}, {\tt executed}, and {\tt observed} fields in the {\tt ldQ}).
For the reachability step, numerous mature, automated verification techniques exist to automate this process, including predicate abstraction, dataflow analysis, and SMT solving. 

Notwithstanding leaving automated predicate checking for future endeavors, our paper has shown, for the first time, a verification approach that is scalable in both the numbers of in-flight memory instructions and of cores in the system. Further, we have automated QED's first step of generating the exploration trees and decision trees of predicates. We believe that this paper makes significant progress toward automatic and scalable verification of hardware consistency. Automating the remaining step is feasible using the current tools. 
\putsec{related}{Related Work}

Sequential Consistency (SC)~\cite{lamport} is the most intuitive memory model and  is implemented in the SGI Origin 2000~\cite{origin2k}. Targeting higher performance,  other models, such as Total Store Order (TSO), Release Consistency (RC) and Relaxed Memory Ordering (RMO), relax various ordering and write atomicity constraints of SC~\cite{tutorial,synth-lecture}. Almost all current, commercially-important CPU families (e.g., x86-64, ARM, Power, and RISC-V) each support a specific such relaxed model. Formal specification of such models is a well-studied topic~\cite{sewell_x86,sewell_arm,sewell_power,sewell_riscv}.
QED's verification remains scalable for all the models. 

Compilers can exploit global knowledge to reorder memory accesses in one thread without affecting the other threads~\cite{snir_toplas88}.
However, this work eliminates unnecessary fences for improved performance; not for MCM verification. 
An early software work~\cite{testing_shared_mem}, attempting to verify  whether a {\em given execution} is sequentially consistent,
shows that the problem of finding the store that provides
the value for a load without violating SC is NP-Complete. 
This approach has been extended for TSO~\cite{tsotool} producing
a partial solution~\cite{complete_verify}. Subsequent works use heuristics to simplify the complexity of the algorithm~\cite{lcheck,xcheck,verify_weakmm}. However, such solutions still do not scale.
In contrast, QED targets scalable verification for hardware consistency. 
As discussed in~\secref{intro}, the early ``*check'' papers~\cite{pipecheck,ccicheck,coatcheck,rtlcheck,tricheck,ilamcm,realitycheck} exhaustively check at the microarchitectural level all executions of a suite of ``litmus tests''. However, there may still be bugs not exposed by the tests~\cite{rtlcheck,pipeproof,comprehensive_asplos17,compare_mcms}. Targeting exhaustive tests, a later paper~\cite{comprehensive_asplos17} generates
comprehensive yet minimal tests with a bounded number of instructions across all threads. Unfortunately, the approach  does not scale in practice to cover modern instruction window sizes. The ``*check'' papers can leverage the exhaustive tests to achieve bounded verification, including that of the  load-store queue~\cite{coatcheck}  and coherence interface~\cite{ccicheck}. Others have enhanced litmus test generation to improve test coverage~\cite{compare_mcms,mcversi,transform}. 
In contrast, PipeProof~\cite{pipeproof} targets unbounded verification by rigorously formulating the problem as a SAT instance.  Kami~\cite{kami} proposes a Bluespec-based, rigorous,   modular approach to tackle verification scalability. PipeProof and Kami verify in-order-issue processors which are much  simpler than modern out-of-order-issue processors.
In contrast, QED focuses on scalably verifying the load-store queue and coherence interface in an out-of-order-issue processor.  
A recent work~\cite{uarchsyn} produces microarchitecture abstractions from  RTL implementations. 

\putsec{conclu}{Conclusion}
To address hardware memory consistency  design bugs, 
we proposed QED, a scalable verification approach, which focuses on the memory ordering issues in an out-of-order processor's load-store queue and the coherence interface between the core 
and global coherence. QED assumes the pipeline front-end register and control-flow dependencies and global coherence (i.e., write serialization, and if required by the MCM, write atomicity) are implemented correctly. QED is based on the key notion of {\em observability} that the memory consistency model (MCM) is violated  only if hardware reordering may produce a forbidden value. 
We argue that (1) only
{\em directly-ordered} instruction pairs -- transitively non-redundant pairs connected by an edge in the MCM-imposed partial  order -- and not all in-flight memory instructions,  and (2) only the ordering of external events from other cores  (e.g., invalidations) but not the events' originating cores.need to be considered. 
Thus, QED 
achieves verification scalability in  both the  numbers of 
in-flight memory instructions and of cores. 
QED exhaustively considers all pairs of instruction types and all types of external events intervening between each pair, 
and attempts to {\em restore} any re-ordered instructions to an MCM-complaint order without changing the execution values (i.e., while remaining unobservable). A failed restoration indicates an MCM violation. 
Each instruction pair's exploration 
gives rise to a decision tree of 
simple, narrowly-defined predicates to be evaluated against the RTL implementation.
In our experiments, we automatically generated the decision trees for SC, TSO, and RISC-V WMO, and  illustrated automatable verification by evaluating  a substantial predicate against  BOOM v3 implementation of RISC-V WMO. Though we leave full automation of predicate evaluation to future work,  QED makes significant progress toward automatic and scalable verification of hardware consistency.

\bibliographystyle{IEEEtranS}
\bibliography{references,coherence_verification}

\end{document}